\documentclass{article}
\usepackage{spconf,amsmath,graphicx,hyperref}
\usepackage{multirow}
\usepackage{graphicx}
\usepackage{booktabs}
\usepackage{xcolor}
\usepackage{cite}
\usepackage{enumitem}

\providecommand{\yw}[1]{\textcolor{black}{{#1}}}

\title{PromptSep: Generative Audio Separation via Multimodal Prompting}
\name{
\parbox{\linewidth}{\centering
Yutong Wen$^{1,2}$\sthanks{Work partly done during an internship at Adobe Research.}, Ke Chen$^{1}$, Prem Seetharaman$^{1}$, Oriol Nieto$^{1}$, Jiaqi Su$^{1}$, Rithesh Kumar$^{1}$ \\ Minje Kim$^{2}$, Paris Smaragdis$^{3}$, Zeyu Jin$^{1}$, Justin Salamon$^{1}$
}}
\address{
         $^1$Adobe Research, USA ~~ $^2$University of Illinois Urbana-Champaign, USA ~~ $^3$MIT, USA
}

\begin{document}
\ninept 

%
\maketitle
\begin{abstract}
\vspace{-0.1cm}
Recent breakthroughs in language-queried audio source separation (LASS) have shown that generative models can achieve higher separation audio quality than traditional masking-based approaches. However, two key limitations restrict their practical use: (1) users often require operations beyond separation, such as sound removal; and (2) relying solely on text prompts can be unintuitive for specifying sound sources. In this paper, we propose PromptSep to extend LASS into a broader framework for general-purpose sound separation. PromptSep leverages a conditional diffusion model enhanced with elaborated data simulation to enable both audio extraction and sound removal. To move beyond text-only queries, we incorporate vocal imitation as an additional and more intuitive conditioning modality for our model, by incorporating Sketch2Sound as a data augmentation strategy. Both objective and subjective evaluations on multiple benchmarks demonstrate that PromptSep achieves state-of-the-art performance in sound removal and vocal-imitation-guided source separation, while maintaining competitive results on language-queried source separation.
\end{abstract}
\vspace{-0.3cm}
\section{Introduction}
\label{sec:intro}
\vspace{-0.2cm}
Audio source separation aims to isolate specific sounds from an audio mixture. 
Prior work has approached this task across various domains, where task-specific sound sources needed to be defined, including speech~\cite{luo2019conv, wang2018supervised}, music~\cite{defossez2019music, rouard2023hybrid}, and general sound events~\cite{ochiai2020listen, kong2023universal}. Target source extraction (TSE) is a variant where a specific source is specified by the users in the form of a conditioning signal for the model, such as class labels~\cite{delcroix2021few, chen2024mdx}, reference audio~\cite{chen2022zero, seetharaman2019class}, visual cues~\cite{li2023target, gao2019co}, and other modalities~\cite{smaragdis2009user, bryan2014isse}. With recent breakthroughs in machine learning, language-queried audio source separation (LASS) has gradually emerged, where natural language serves as the conditioning input for target audio source separation~\cite{dong2022clipsep, liu2022separate, liu2024separate, yuan2025flowsep, wang2025soloaudio, hai2024dpm}.  

Conventional source separation methods have largely been dominated by mask prediction models that minimize point-wise losses between masked audio and target audio signals~\cite{luo2019conv, hsieh2025tgif}. However, masking-based models often introduce distortions, artifacts, and leakages in separation results, as maintaining mask consistency on original audio becomes particularly challenging when multiple sounds overlap. Recently, generative models like diffusion and flow-matching are alternatives to the masking-based methods for separation tasks~\cite{zhu2025review, zhu2022music, mariani2023multi, subakan2018generative, wen2025user}, achieving higher separation audio quality. And such methods are also adopted in the LASS settings~\cite{hai2024dpm, wang2025soloaudio}.

Despite promising progress, LASS systems still face two major limitations in real-world scenarios. First, most audio separation models only ``extract" the target source from audio mixtures, while
treating separation solely as an extraction operator is restrictive. Users may also wish to remove specific sounds from an audio mixture (i.e., removal), rather than only isolating specific sounds (i.e., extraction). Supporting both extraction and removal within a single framework would better align with real-world needs. Masking-based approaches, 
however, struggle to generalize to such multi-operator use cases, as they often fail to deliver an accurate one-time mask for multiple sound targets for removal purpose~\cite{kongWSCWP20, chen2022zero, kong2023universal}. In contrast, generative approaches offer an alternative. It is worth exploring whether generative models can synthesize high-quality audio outputs that support both extraction and removal operators by explicitly modeling the data distribution of high-fidelity audio samples.

Second, language may not be the most appropriate and effective query for audio sources. Some textual descriptions of sounds, such as ``distortion", ``dark transition", or ``light flashing", are too abstract or ambiguous to precisely specify the target sources. Additionally, a single sound can be described in many different ways, leading to high variability in prompt length, vocabulary, and phrasing. Multiple sources in a mixture may also satisfy a given description (e.g., ``distortion" could refer to any harsh or intense sounds), causing unwanted sources to be extracted or removed. Most importantly, this limitation is inherent to language itself, as sounds are naturally perceived by human ears rather than captured by textual descriptions. To overcome this limitation, previous work~\cite{chen2022zero} has explored audio prompts, example samples indicating the target source, to separate similar sounds in the mixture. But this approach is also unintuitive, as users must still provide appropriate reference audio samples.

To address the above limitations, we propose PromptSep, a latent diffusion model for open-vocabulary target audio source separation with multimodal cues. 
Specifically, we first extend the standard extraction-only separation process with \textbf{removal} operator.
Second, we incorporate \textbf{vocal imitation} as another conditioning signal beyond text prompts, where users can mimic the target sound to guide the separation target. This makes the system align better with how humans naturally describe sounds. To achieve this, we explore how to augment vocal imitation data by leveraging existing sound effect generation models~\cite{garcia2025sketch2sound}, datasets~\cite{kim2019vim}, and modules~\cite{wen2025user}. With PromptSep, users can perform either sound extraction or removal by leveraging either a textual description or a vocal imitation as a query. Vocal imitation alleviates the need to locate audio samples when the textual description is ambiguous. 
Our contributions are three-fold: 
\vspace{-0.1cm}
\begin{enumerate}[leftmargin=*]
    \item We extend extraction-only source separation with a conditional diffusion model and data simulation pipeline to support sound removal, offering more flexible separation operators.
    \vspace{-0.1cm}
    \item We incorporate vocal imitation as an additional query modality via data augmentation and conditional modules, offering a more intuitive source control than text.
    \vspace{-0.1cm}
    \item We provide a thorough evaluation and demonstrate PromptSep achieves superior performance on sound removal and vocal-imitation-guided extraction, and maintains competitive results on standard language-queried separation. 
\end{enumerate}

\section{Method}
\label{sec:method}
\vspace{-0.1cm}

\begin{figure}[t]
  \centering
  \includegraphics[width=\columnwidth]{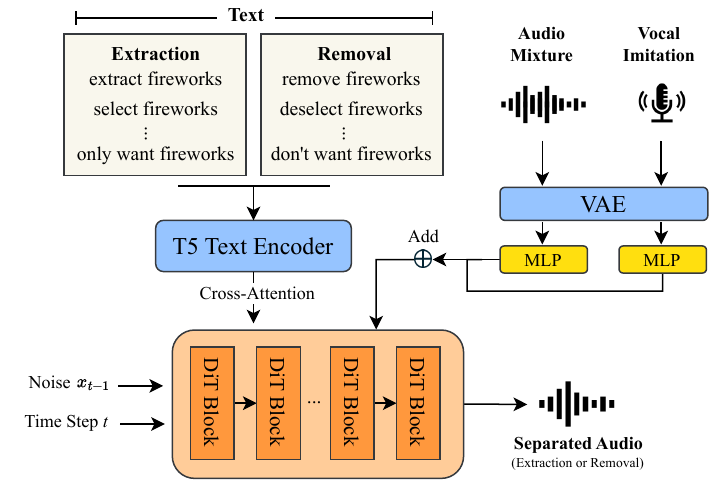}
  \caption{The model architecture of PromptSep. Text and vocal imitation inputs can be used separately or combined.}
  \label{fig:overview}
  \vspace{-0.4cm}
\end{figure}

As shown in Figure~\ref{fig:overview}, PromptSep takes an audio mixture as the primary input, along with two conditions: (1) a textual description of the separation target; and (2) a vocal imitation recording of the separation target. Both conditions can be used separately or jointly. Based on these inputs, the model outputs an audio track containing the sources specified by the conditions.  In the following subsections, we introduce the construction of these conditions and the model architecture of PromptSep.

\vspace{-0.25cm}
\subsection{Separation Condition}
\vspace{-0.1cm}

\textbf{Text Prompt}\; The textual description of sounds varies in complexity. They may consist of just one keyword identifying the sound, or a long sentence containing multiple sound attributes. Consequently, text prompts can range from a single word to several sentences. To prepare the model for such a condition, we include text prompts of different lengths and descriptive styles for the same sounds during training, simulating the user inputs in real-world scenarios.

Moreover, we extend support for both the number of target sounds and separation operations. Previous LASS approaches typically rely on text prompts that specify a single target sound, and the models are also designed to handle only one target sound at a time. This constraint reduces practicality in real-world scenarios, where users want to extract multiple sounds simultaneously. In addition, users sometimes want to remove specific sounds rather than extract them. Indeed, such removal cases are particularly common, as it is often easier to describe the sounds to remove while preserving the remaining sounds. Previous approaches do not support this sort of interactions, to the best of our knowledge. 

To address this gap, we first generalize the target of a single sound source to any subset of sources in the mixture. 
Then we introduce two text-based operators, \textbf{extraction} and \textbf{removal}, and use GPT-5 to generate 100 textual templates that paraphrase extraction or removal operations (50 for each). 
The specification of these templates can be viewed at our accompaniment  webpage\footnote{\href{https://yutongwen.github.io/PromptSep/}{https://yutongwen.github.io/PromptSep/}}. 

We conduct the data simulation pipeline to combine these templates with captions of the target sounds to generate text prompts of either extraction or removal operations. We combine them with corresponding input mixtures and output targets for the language-queried audio separation training.  


\vspace{0.2cm}

\noindent\textbf{Vocal Imitation}\; As mentioned in Section~\ref{sec:intro}, certain sounds (e.g., ``distortion" and ``buzzing") can be too abstract to describe accurately using text alone, or too ambiguous for the model to only identify the single candidate. To enable more flexible and intuitive sound specification, we introduce an additional guidance modality for separation: vocal imitation. PromptSep can be conditioned on reference audio samples in which a user vocally imitates a target sound, thereby guiding the separation process. This approach provides a more natural and accessible way for users to describe the sounds.

While a few datasets, such as VimSketch~\cite{kim2019vim}, contain pairs of corresponding vocal imitation and sound effect samples, 
these pairs are not temporally aligned, and thus cannot serve as a good vocal imitation condition to identify the target sound in a mixture.

To address this, we leverage the sound effect generation model Sketch2Sound~\cite{garcia2025sketch2sound} for data augmentation. 
It can generate sound effect audio samples that match textual descriptions and are temporally aligned with vocal imitation prompts.
Using around 12K real vocal imitation samples and their corresponding sound labels from the VimSketch dataset, we use Sketch2Sound to generate around 87K temporally aligned sound effect samples as training data for PromptSep. To better simulate real-world conditions, we apply time-shift and pitch-shift augmentations as imitation variances. We also add ambient noise, from 4.36 hours of static noise data collections~\cite{kinoshita2013reverb, eaton2016estimation,chen2021structure}, to enhance the model generalization capability.

We note that the Sketch2Sound training data does not require paired vocal imitations and sound effects. 
Instead, it relies on RMS and pitch curves to guide sound generation, without using actual imitation samples. 
This leads to a possible exploration: whether curve-based features (RMS and pitch) or actual vocal-imitation raw waveform inputs provide more effective conditioning for audio separation. In Section~\ref{sec:results}, we address this through ablation studies.

\vspace{-0.25cm}
\subsection{Model Architecture}
\vspace{-0.1cm}

We implement PromptSep using a latent diffusion model via diffusion Transformer (DiT) with three input types (text, vocal imitation, and audio mixture), following the specifications of~\cite{garcia2025sketch2sound, evans2024fast, evans2024long}. 
It contains three main components: (1) a pretrained variational autoencoder (VAE), following the architecture of Descript Audio Codec (DAC)~\cite{kumar2023high}, compresses 44.1 kHz mono audio samples into a sequence of continuous 128-dimensional embeddings at a temporal resolution of 40 Hz; (2) a pretrained FLAN-T5 encoder~\cite{raffel2019exploring} encodes all text prompts; and (3) a DiT model is trained to generate new sequence of embeddings, which are decoded back to waveform via the VAE decoder to reconstruct target separated audio.

For conditioning both the mixture and the vocal imitation samples, 
we adopt an in-place addition mechanism following~\cite{garcia2025sketch2sound, wu2024music, wen2025user} as they have the same latent dimensionality as the separated audio. Specifically, we apply a single MLP layer respectively to mixture and vocal imitation, and add the resulting embeddings to the noisy latent as input to DiT.

As we allow the sum of multiple sounds as a target, a trivial solution to the learning objective is to replicate the input mixture to achieve a deceptively low loss. A light noise perturbation to the input can prevent this case, as replicating won't achieve low loss values.

\begin{table*}[t]
\centering
\resizebox{\textwidth}{!}{
\begin{tabular}{lcccccccccccccccccc}
\toprule
\multirow{2}{*}{Models} &
  \multicolumn{3}{c}{SDRi $\uparrow$} &
  \multicolumn{3}{c}{L2 Mel $\downarrow$} &
  \multicolumn{3}{c}{F1 Decision Error $\uparrow$} &
  \multicolumn{3}{c}{CLAPScore $\uparrow$} &
  \multicolumn{3}{c}{$\text{CLAPScore}_{A} \uparrow$} &
  \multicolumn{3}{c}{FAD$_{\textit{PANN}}$ $\downarrow$} \\ \cmidrule{2-19} 
 &
  ACESC &
  FSD &
  \multicolumn{1}{c|}{ASFX} &
  ACESC &
  FSD &
  \multicolumn{1}{c|}{ASFX} &
  ACESC &
  FSD &
  \multicolumn{1}{c|}{ASFX} &
  ACESC &
  FSD &
  \multicolumn{1}{c|}{ASFX} &
  ACESC &
  FSD &
  \multicolumn{1}{c|}{ASFX} &
  ACESC &
  FSD &
  ASFX \\ \midrule
FlowSep~\cite{yuan2025flowsep} &
  -4.26 &
  2.05 &
  \multicolumn{1}{c|}{-2.75} &
  \textbf{3.06} &
  13.80 &
  \multicolumn{1}{c|}{4.93} &
  \textbf{0.88} &
  0.45 &
  \multicolumn{1}{c|}{0.55} &
  0.24 &
  0.11 &
  \multicolumn{1}{c|}{0.14} &
  \textbf{0.74} &
  0.45 &
  \multicolumn{1}{c|}{0.63} &
  23.70 &
  50.39 &
  38.68 \\
SoloAudio~\cite{wang2025soloaudio} &
  \textbf{2.42} &
  \textbf{14.75} &
  \multicolumn{1}{c|}{5.15} &
  8.35 &
  \textbf{2.26} &
  \multicolumn{1}{c|}{4.73} &
  0.74 &
  \textbf{0.80} &
  \multicolumn{1}{c|}{0.53} &
  \textbf{0.26} &
  \textbf{0.30} &
  \multicolumn{1}{c|}{\textbf{0.19}} &
  0.62 &
  \textbf{0.77} &
  \multicolumn{1}{c|}{0.61} &
  21.17 &
  \textbf{5.79} &
  8.82 \\
PromptSep &
  1.74 &
  10.89 &
  \multicolumn{1}{c|}{\textbf{5.65}} &
  5.04 &
  7.60 &
  \multicolumn{1}{c|}{\textbf{4.23}} &
  0.82 &
  0.60 &
  \multicolumn{1}{c|}{\textbf{0.60}} &
  0.24 &
  0.22 &
  \multicolumn{1}{c|}{0.18} &
  0.62 &
  0.58 &
  \multicolumn{1}{c|}{\textbf{0.66}} &
  \textbf{12.00} &
  19.75 &
  \textbf{3.19} \\ \bottomrule
\end{tabular}
}
\vspace{-0.3cm}
\caption{Results under the standard extraction setup, evaluated on AudioCaps + ESC50 (ACESC)~\cite{yuan2025flowsep}, FreeSound (FSD)~\cite{wang2025soloaudio}, and Adobe Audition Sound Effects (ASFX). Among three benchmarks, ASFX is the only evaluation set that is out-of-domain for all three models.}
\label{tab:extract}
\vspace{-0.2cm}
\end{table*}

\begin{table*}[t]
\centering
\resizebox{\textwidth}{!}{
\begin{tabular}{lccccccccccccccc}
\toprule
 &
  \multicolumn{3}{c}{SDRi $\uparrow$} &
  \multicolumn{3}{c}{L2 Mel $\downarrow$} &
  \multicolumn{3}{c}{F1 Decision Error $\uparrow$} &
  \multicolumn{3}{c}{$\text{CLAPScore}_{A} \uparrow$} &
  \multicolumn{3}{c}{$\text{FAD}_{\textit{PANN}} \downarrow$} \\ \cmidrule{2-16} 
\multirow{-2}{*}{Models} &
  ACESC &
  FSD &
  \multicolumn{1}{c|}{ASFX} &
  ACESC &
  FSD &
  \multicolumn{1}{c|}{ASFX} &
  ACESC &
  FSD &
  \multicolumn{1}{c|}{ASFX} &
  ACESC &
  FSD &
  \multicolumn{1}{c|}{ASFX} &
  ACESC &
  FSD &
  ASFX \\ \midrule
FlowSep~\cite{yuan2025flowsep} &
  -4.45 &
  -12.44 &
  \multicolumn{1}{c|}{-9.53} &
  \textbf{6.30} &
  13.27 &
  \multicolumn{1}{c|}{5.99} &
  0.76 &
  0.70 &
  \multicolumn{1}{c|}{0.56} &
  0.44 &
  0.55 &
  \multicolumn{1}{c|}{0.61} &
  25.49 &
  28.74 &
  20.26 \\
SoloAudio~\cite{wang2025soloaudio} &
  -1.08 &
  -10.85 &
  \multicolumn{1}{c|}{-5.50} &
  12.40 &
  37.84 &
  \multicolumn{1}{c|}{10.87} &
  0.59 &
  0.29 &
  \multicolumn{1}{c|}{0.36} &
  0.30 &
  0.20 &
  \multicolumn{1}{c|}{0.45} &
  27.20 &
  87.79 &
  18.54 \\
PromptSep &
  \textbf{1.17} &
  \textbf{-3.34} &
  \multicolumn{1}{c|}{\textbf{-3.20}} &
  6.40 &
  \textbf{9.13} &
  \multicolumn{1}{c|}{\textbf{4.86}} &
  \textbf{0.80} &
  \textbf{0.87} &
  \multicolumn{1}{c|}{\textbf{0.75}} &
  \textbf{0.54} &
  \textbf{0.71} &
  \multicolumn{1}{c|}{\textbf{0.72}} &
  \textbf{16.27} &
  \textbf{15.99} &
  \textbf{3.81} \\ \midrule
{\color[HTML]{9B9B9B} FlowSep*~\cite{yuan2025flowsep}} &
  {\color[HTML]{9B9B9B} -4.35} &
  {\color[HTML]{9B9B9B} -13.14} &
  \multicolumn{1}{c|}{{\color[HTML]{9B9B9B} -9.36}} &
  {\color[HTML]{9B9B9B} 3.01} &
  {\color[HTML]{9B9B9B} 6.64} &
  \multicolumn{1}{c|}{{\color[HTML]{9B9B9B} 3.34}} &
  {\color[HTML]{9B9B9B} 0.88} &
  {\color[HTML]{9B9B9B} 0.87} &
  \multicolumn{1}{c|}{{\color[HTML]{9B9B9B} 0.73}} &
  {\color[HTML]{9B9B9B} 0.74} &
  {\color[HTML]{9B9B9B} 0.76} &
  \multicolumn{1}{c|}{{\color[HTML]{9B9B9B} 0.74}} &
  {\color[HTML]{9B9B9B} 24.37} &
  {\color[HTML]{9B9B9B} 13.37} &
  {\color[HTML]{9B9B9B} 19.78} \\
{\color[HTML]{9B9B9B} SoloAudio*~\cite{wang2025soloaudio}} &
  {\color[HTML]{9B9B9B} 2.26} &
  {\color[HTML]{9B9B9B} -9.82} &
  \multicolumn{1}{c|}{{\color[HTML]{9B9B9B} -3.77}} &
  {\color[HTML]{9B9B9B} 8.60} &
  {\color[HTML]{9B9B9B} 35.31} &
  \multicolumn{1}{c|}{{\color[HTML]{9B9B9B} 8.70}} &
  {\color[HTML]{9B9B9B} 0.74} &
  {\color[HTML]{9B9B9B} 0.44} &
  \multicolumn{1}{c|}{{\color[HTML]{9B9B9B} 0.54}} &
  {\color[HTML]{9B9B9B} 0.62} &
  {\color[HTML]{9B9B9B} 0.32} &
  \multicolumn{1}{c|}{{\color[HTML]{9B9B9B} 0.59}} &
  {\color[HTML]{9B9B9B} 23.57} &
  {\color[HTML]{9B9B9B} 78.45} &
  {\color[HTML]{9B9B9B} 14.20} \\
\bottomrule
\end{tabular}
}
\vspace{-0.3cm}
\caption{Results under the removal setup using negative text operators. We include a ``upper-anchor" setup where models achieve the same removal effect by separating multiple target sounds as an upper bound of previous baselines (marked with * and in gray).}
\label{tab:neg}
\vspace{-0.3cm}
\end{table*}

During training, we randomly drop condition signals for classifier-free guidance (CFG), with the drop rate $10\%$ for text and mixture, but $90\%$ for vocal imitation, as it gets easier to overfit through our experiments.  We use a $v$-prediction framework for training and diffusion probabilistic models (DPM) solvers~\cite{lu2025dpm} for sampling. During inference, we set the CFG scale to 1.0.

\section{Experiments}
\vspace{-0.25cm}
\subsection{Training Datasets}

\vspace{-0.1cm}
\textbf{Sound Event}\; We train our model using an internal large collection of licensed sound effect datasets and publicly available, CC-licensed general audio corpora. Each sound is accompanied with multiple versions of captions varying in length.
As described in Section~\ref{sec:method}, we combine the captions with the text operator templates to form the final text input. \yw{Figure~\ref{fig:overview} shows example final text input.}
\vspace{0.15cm}

\noindent\textbf{Vocal Imitation}\; We curate a new dataset, VimSketchGen, consisting of $87,171$ pairs of aligned vocal imitations and sound effects. 
This dataset originally contains $12,453$ vocal imitations sourced from the VimSketch dataset, and each of them is paired with $7$ corresponding sound effects generated using Sketch2Sound, with different median filter sizes $\in \{0, 3, 6, 9, 12, 15, 19\}$. 
All audio samples in VimSketchGen are 8-second stereo tracks sampled at 44.1 kHz. We will release this dataset to support more tasks in audio research.
\vspace{0.15cm}

\noindent\textbf{Training Specification}\; The input of PromptSep is a 10-second audio mixture by randomly combining $2$ to $5$ sound events from different categories
. The mixing signal-to-noise ratio (SNR) is uniformly sampled between $-3$ and $10$ dB. A random subset of sound events in the mixture is selected as the separation target, with the only exception that if the vocal imitation is chosen as condition, its corresponding sound event alone is used as the target. 
\yw{During training, the model is always conditioned on either text or vocal imitation, but not both simultaneously. While both conditions can be provided at inference time, their combined effect is not explored in this work and is left for future investigation.
}



\vspace{-0.15cm}
\subsection{Evaluation Datasets}
\noindent\textbf{AudioCaps~\cite{kim2019audiocaps}}\; We follow the evaluation setup from FlowSep\cite{yuan2025flowsep} to use the AudioCaps test set of $928$ audio clips. 
We treat each audio clip as the target source and mix it with a noise clip randomly selected from the test set at an SNR randomly chosen between $-15$ dB and $15$ dB. 
The first caption associated with the target audio is used as the query for separation.
\vspace{0.15cm}

\noindent\textbf{ESC50~\cite{piczak2015esc}}\; It is also from the evaluation setup in FlowSep. We use the ESC50 evaluation set of $2000$ audio clips. Similarly, each clip is mixed with another randomly selected clip at an SNR of $0$ dB.
\vspace{0.15cm}

\noindent\textbf{FSD-Mix~\cite{wang2025soloaudio}}\; We use the test set of FSD-Mix, which contains $1,440$ audio mixtures. 
Each mixture consists of $3$ to $5$ sound events, mixed at an SNR randomly selected between $-10$ and $10$ dB.
\vspace{0.15cm}

\noindent\textbf{Adobe Audition Sound Effects~\footnote{\href{https://www.adobe.com/products/audition/offers/adobeauditiondlcsfx.html}{https://www.adobe.com/products/adobeauditiondlcsfx}}}\; It serves as a completely out-of-domain benchmark, as both PromptSep and previous baselines are not trained on any training, validation, and test sets of it. We create $2000$ mixtures and each contains $2$ to $5$ sound events randomly sampled from the test set, with an SNR between $-10$ and $10$ dB.
\vspace{0.15cm}

\noindent\textbf{VimSketchGen-Mix}\; We use a subset of the VimSketchGen test split, containing $2000$ sound event samples with their vocal imitation pairs. Each target sound is mixed with 1 to 3 interference sounds randomly selected from the AudioSet test set~\cite{gemmeke2017audio}, using a SNR sampled between $-3$ and $10$ dB.

\vspace{-0.15cm}
\subsection{Baselines}
We compare our model against two generative language-queried audio separation models: FlowSep~\cite{yuan2025flowsep} and SoloAudio~\cite{wang2025soloaudio}. 
FlowSep performs separation in the latent space of a Mel-spectrogram VAE using flow matching, and reconstructs audio using a vocoder. 
It employs FLAN-T5 for text conditioning. 
SoloAudio, on the other hand, applies a modified DiT architecture in the VAE latent space and uses CLAP~\cite{wu2023large} for text embeddings. 

\vspace{-0.15cm}
\subsection{Objective Metrics}

We evaluate the performance of separation models using 6 objective metrics. Signal-to-distortion ratio improvement (SDRi) and L2 multi-resolution Mel-spectrogram distance~\cite{kumar2023high} are the most conventional metrics to measure the signal-level differences between the output and the groundtruth. Following~\cite{xiao2024reference, yuan2025flowsep,wang2025soloaudio}, we also include CLAPScore, CLAPScore$_{A}$, and Frechet Audio Distance (FAD)~\cite{kilgour2018fr} with embeddings from PANNs~\cite{kong2020panns}, to assess the generation audio quality and the semantic correlation among the text prompt, separation audio, and groundtruth audio.


To better evaluate the separation decision error, we propose \textbf{F1 Decision Error} as a new metric to evaluate the ability of the model to identify the correct temporal regions of a target sound. Specifically, to obtain the F1 score of decision errors, we first compute the frame-wise RMS energy on both the separated audio and the groudtruth audio. These values are then binarized using a threshold (0.01) to obtain the activity sequences, determining the sound and unsound frames. Finally, we calculate the F1 score between the predicted activity sequence and the groundtruth activity sequence.

\begin{table}[t]
\centering
\resizebox{\columnwidth}{!}{
\begin{tabular}{lcccc}
\toprule
\multirow{2}{*}{Model} & \multicolumn{2}{c}{Extraction} & \multicolumn{2}{c}{Removal} \\ \cmidrule{2-5} 
 & REL$\uparrow$ & OVL$\uparrow$ & REL$\uparrow$ & OVL$\uparrow$ \\ \midrule
Mixture & 2.96 $\pm$ 0.08 & 3.55 $\pm$ 0.07 & 2.42 $\pm$ 0.08 & 3.26 $\pm$ 0.08 \\
GT & 3.94 $\pm$ 0.07 & 4.17 $\pm$ 0.06 & 3.27 $\pm$ 0.08 & 4.06 $\pm$ 0.06 \\ \midrule
FlowSep~\cite{yuan2025flowsep} & 3.19 $\pm$ 0.07 & 3.46 $\pm$ 0.07 & 2.88 $\pm$ 0.08 & 3.40 $\pm$ 0.07 \\
SoloAudio~\cite{wang2025soloaudio} & 3.31 $\pm$ 0.08 & 3.64 $\pm$ 0.07 & 2.99 $\pm$ 0.09 & 3.59 $\pm$ 0.07 \\
PromptSep & \textbf{3.34 $\pm$ 0.08} & \textbf{3.75 $\pm$ 0.07} & \textbf{3.25 $\pm$ 0.08} & \textbf{3.83 $\pm$ 0.07} \\ \bottomrule
\end{tabular}}
\vspace{-0.2cm}
\caption{Mean Opinion Scores (MOS) with standard error for text relevance (REL) and overall quality (OVL). Mixture and GroudTruth (GT) serve as the lower-anchor and upper-anchor results.}
\label{tab:subj_test}
\vspace{-0.3cm}
\end{table}

\section{Results} \label{sec:results}
\subsection{Language-queried Target Sound Extraction}

Table~\ref{tab:extract} presents the results for the language-queried target sound extraction setup. Each model is provided with a text description of the target sounds and evaluated on its ability to extract the sounds from a mixture. We compare PromptSep against two baselines on four benchmarks. Due to the page limitation, we aggregate results from AudioCaps and ESC50 as ACESC using a weighted sum over their metrics, proportional to the number of samples in each dataset.

PromptSep achieves the best performance on nearly all metrics on ASFX, with the exception of CLAPScore, where SoloAudio surpasses us by 0.01. This highlights the strong generalization of our model, as ASFX is the only out-of-domain test set for all three models. FlowSep, trained on AudioSet, WavCaps, and VGGSound, which share similar quality and sources with AudioCaps and ESC50, obtains the best L2 Mel distance, F1 decision error, and CLAPScore${A}$ on ACESC. SoloAudio, trained primarily on FreeSound, achieves the best results on FSD. All of AudioCaps, ESC50, and FSD are out-of-domain for PromptSep, but it yields competitive performance: achieving or nearly matching the best scores in FAD, SDRi, F1, CLAPScore, and CLAPScore${A}$ on ACESC, and in SDRi on FSD. These results further demonstrate the strong generalization ability of PromptSep.

Finally, we note that FlowSep performs poorly on SDRi. This is likely due to its separation process, which relies on vocoder to reconstruct the generated mel-spectrograms. While its output may preserve acoustic patterns and types of sound events, it can deviate substantially from the ground-truth waveform at the signal level.

\vspace{-0.2cm}
\subsection{Language-queried Target Sound Removal}

Beyond the standard extraction setup, PromptSep also supports the sound removal. While our baselines were not explicitly trained with removal operations, they have been exposed to large-scale textual descriptions and may have implicitly learned some removal capability. Therefore, we also include their results for comparison. 
As presented in Table~\ref{tab:neg}, PromptSep outperforms the baselines across all datasets and metrics, with the exception of the L2 Mel distance on ACESC. These highlight the strong performance of our model in language-queried target sound removal.

To further account for the limitations of baselines on sound removal, we design an alternative setup. Instead of directly removing the target sound, models are prompted to extract the remaining sounds by providing combined text descriptions of all non-target events, as an equivalent operation. Results under this configuration are shown in Table~\ref{tab:neg} (gray rows, with models marked by *). PromptSep continues to achieve most of the best scores, even compared against FlowSep* and SoloAudio*. This further demonstrates the strong performance of our model in sound removal.

\begin{table}[]
\centering
\resizebox{\columnwidth}{!}{
\begin{tabular}{lccccc}
\toprule
Conditions & SDRi $\uparrow$           & L2 Mel $\downarrow$       & F1 Decision Error $\uparrow$           & $\text{CLAPScore}_{A} \uparrow $& FAD $\downarrow$           \\ \midrule
Imitation  & \textbf{9.99} & \textbf{0.92} & \textbf{0.95} & \textbf{0.87}   & \textbf{2.19} \\
Pitch+RMS  & 7.17          & 3.30          & 0.84          & 0.71            & 6.66          \\
\bottomrule
\end{tabular}}
\vspace{-0.2cm}
\caption{Results for the vocal imitation (Imitation) condition, along with the Pitch and RMS condition for ablation analysis.}
\label{tab:vim}
\vspace{-0.3cm}
\end{table}

\subsection{Subjective Evaluation}
We further conduct a subjective evaluation for both the extraction and removal setups by following the format of DCASE2024 Task 9~\footnote{\href{https://dcase.community/challenge2024/task-language-queried-audio-source-separation}{https://dcase.community/challenge2024/task-language-queried-audio-source-separation}}, using Relevance (REL) and Overall Sound Quality (OVL). The REL score measures how well the separated audio matches the given language query. 
The OVL score evaluates the perceived audio quality of the output, including factors such as clarity, naturalness, and artifacts. 
Both REL and OVL are rated by human annotators using a 5-point Likert scale.

We randomly select 100 samples from ASFX test set as it is the only out-of-domain benchmark for all three models. The same mixtures and text descriptions are used across both the extraction and removal setups to ensure consistency. The total number of participants is 100, with each of samples are rated by at least 4 participants.
Results are shown in Table~\ref{tab:subj_test}. 
PromptSep achieves the highest scores in both REL and OVL across both setups, demonstrating its strong performance in accurately separating target sounds and maintaining high audio quality.

\vspace{-0.2cm}
\subsection{Imitation-queried Target Sound Extraction}

To the best of our knowledge, no existing system supports vocal imitations as a conditioning input for open-domain source separation. 
We evaluate our model on the VimSketchGen-Mix with no baseline. 
Results are presented in Table~\ref{tab:vim}, where our model achieves an SDRi of $9.99$ dB, an L2 multi-resolution Mel-spectrogram distance of $0.92$, an F1 Decision Error of $0.95$, a CLAPScore$_\text{A}$ of $0.87$, and a FAD score of $2.19$.
These results indicate strong separation performance and demonstrate that vocal imitation is an effective conditioning signal for source separation.

We also evaluate a variant of our model that uses only pitch and RMS features extracted from the vocal imitation as the conditioning input, this setup is trained using the same median filter strategy as in \cite{garcia2025sketch2sound}, with the median filter size fixed to 8 during inference. While the pitch and RMS-based conditioning yields reasonably strong separation performance, it consistently under-performs the full vocal imitation condition across all evaluation metrics. 
We attribute this to the complexity of the mixtures, where overlapping sounds are common; in such cases, the raw vocal imitation provides richer information than the limited pitch and RMS features alone.

\vspace{-0.1cm}
\section{Conclusion}
\vspace{-0.1cm}

PromptSep offers a unified framework for sound extraction and removal that overcomes key limitations of existing LASS systems. 
Our approach supports both sound extraction and removal within a single model. 
Furthermore, the integration of vocal imitation as a query modality addresses the ambiguity and limitations of text prompts, offering a more intuitive interface for users. 
Through comprehensive evaluations, PromptSep demonstrates SOTA performance in sound removal and vocal-imitation-guided separation, while remaining competitive in standard LASS settings. 
\vfill\pagebreak


\footnotesize
\bibliographystyle{IEEEbib}
\bibliography{strings,refs}

\end{document}